\documentstyle[12pt,aps,epsf]{revtex}
\begin{document}
\thispagestyle{empty}
\begin{center}
{\large \bf Spectroscopy of doubly heavy baryons}\\
\vspace*{5mm}
S.S.Gershtein, {V.V.Kiselev}, A.K.Likhoded, A.I.Onishchenko\\

\vspace*{7mm}
{\sf State Research Center "Institute for High Energy Physics"} \\
{\it Protvino, Moscow region, 142284 Russia}\\
Fax: +7-095-2302337\\
E-mail: kiselev@th1.ihep.su
\end{center}
\begin{abstract}
{Spectra of masses are calculated for the families of doubly heavy baryons in
the framework of nonrelativistic quark model with the QCD potential by Buchm\"
uller--Tye. We suppose the quark-diquark structure for the wave functions and
take into account the spin-dependent splittings. The physical reasons causing
the existence of quazi-stable excited states in the subsystem of heavy diquark
are considered for the heavy quarks of identical flavors.}
\end{abstract}

\vspace*{1cm}

\newpage
\setcounter{page}{1}
\section{Introduction}

The first observation of $B_c^+$ meson by the CDF collaboration \cite{1} opens
a new direction in the physics of hadrons containing heavy quarks. This
particle completes the list of heavy quarkonia  $(Q\bar Q')$ and heavy flavored
mesons accessible for the experimental investigations. It begins another list
of long-lived hadrons containing two heavy quarks. So, in addition to the
$B_c^+$ meson this class of hadrons would be continued by the doubly heavy
baryons $\Xi_{cc}$, $\Xi_{bc}$ and $\Xi_{bb}$ (see notations in the framework
of quark model by PDG \cite{2}). The experimental discovery of $B_c^+$ was
prepared by theoretical studies of the meson spectroscopy as well as the
mechanisms of its production and decay (see review in \cite{3}). To observe
the doubly heavy baryons it is necessary to give reliable theoretical
predictions of their properties. The initial steps forward such the goal were
done.
\begin{enumerate}
\item
In ref.\cite{4} the estimates for the lifetimes of $\Xi_{cc}^+$ and
$\Xi_{cc}^{++}$ baryons were obtained in the framework of Operator Product
Expansion in the inverse heavy quark mass.
\item
Papers \cite{5} were devoted to the investigation of differential and
total cross-sections for the production of $\Xi_{QQ'}$ baryons in various
interactions in the model of fragmentation, in the model of intrinsic charm
\cite{6} (for the hadronic production of $\Xi_{cc}$) and in the framework of
perturbative QCD calculations up to $O(\alpha_s^4)$ contributions, taking
into account the hard nonfragmentational regime in addition to the
fragmentation, which dominates at high transverse momenta $p_T\gg M$.
\item
In refs.\cite{7,7a} the masses of ground states of doubly heavy baryons were
estimated, and the excitations of $\Xi_{cc}$ were considered in ref.\cite{8}.
\end{enumerate}

In the present paper we analyze the basic spectroscopic characteristics for the
families of doubly heavy baryons $\Xi_{QQ'}=(QQ'q)$, where $q=u,\; d$ and
$\Omega_{QQ'}=(QQ's)$.

A general approach of potential models to calculate the masses of baryons
containing two heavy quarks was considered in refs.\cite{Wise}. The physical
motivation used the pair interactions between the quarks composing the baryon,
that was explored in the three-body problem. Clear implications for the mass
spectra of doubly heavy baryons were derived. So, for the given masses of heavy
charmed and beauty quarks, the approximation of factorization for the motion of
doubly heavy diquark and light quark is not accurate. It results in the ground
state mass and excitation levels, essentially deviating from the estimates in
the framework of appropriate three-body problem. For example, we can easily
find that in the oscillator potential of pair interactions an evident
introduction of Jacobi variables leads to the change of vibration energy
$\omega\to \sqrt{\frac{3}{2}} \omega$ in comparison with naive expectations
of diquark factorization.

\begin{figure}[th]
\begin{center}
\epsfxsize=10cm \epsfbox{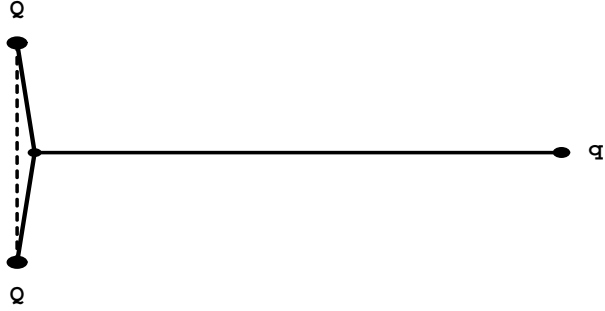}
\end{center}
\caption{The picture of doubly heavy baryon $QQq$ with the colored fields
forming the strings between the heavy and light quarks, that destroys the pair
interactions and involves the additional `centre-of-mass' point close to the
centre of mass for the heavy-heavy system.}
\label{pic-str}
\end{figure}

However, the string-like picture of doubly heavy baryon shown in
Fig.\ref{pic-str}, certainly destroys the above conclusions based on the pair
interactions. Indeed, to the moment we have to introduce the centre of
string, which is very close to the centre of mass for the doubly heavy
diquark. Furthermore, the light quark interacts with the doubly heavy
diquark as a whole, i.e. with the string tension identical to that in the
heavy-light mesons. Therefore, two different assumptions on the nature of
interactions inside the doubly heavy baryons: pair interactions or string-like
picture, result in a distinct variation of predictions on the mass spectra of
these baryons: the ground states and excitation levels. The only criterion
testing the assumptions is provided by an experimental observation and
measurements.

In this paper we follow the approximation of double heavy diquark, which is
quite reasonable as we have clarified in the discussion given above. To enforce
this point we refer to the consideration of doubly heavy baryon masses in the
framework of QCD sum rules \cite{7a}, that recently was essentially improved in
paper \cite{ba-nrqcd}, exploring the NRQCD version of sum rule method. The
conclusion drawn in the sum rules is the ground state mass is in a good
agreement with the estimates obtained in the potential approach with the
factorization of doubly heavy diquark.

The qualitative picture for the forming of bound states in the system of
$(QQ'q)$ is determined by the presence of two scales of distances, which are
given by the size of $QQ'$-diquark subsystem, $r_{QQ'}$, in the anti-triplet
color state  as well as by the confinement scale, $\Lambda_{QCD}$, for the
light quark $q$, so that
$$
r_{QQ'}\cdot \Lambda_{QCD} \ll 1, \;\;\;\; \Lambda_{QCD}\ll m_Q.
$$
Under such conditions, the compact diquark $QQ'$ looks like a static source
approximated by the local colored QCD field interacting with the light quark.
Therefore, we may use a set of reliable results in models of mesons with a
single heavy quark, i.e. with a local static source belonging to the
anti-triplet representation of SU(3)$_c$ group. The successful approaches are
the potential models \cite{9} and the Heavy Quark Effective Theory (HQET)
\cite{10} in the framework of expansion in the inverse heavy quark mass. We
apply the nonrelativistic quark model with the potential by Buchm\"uller--Tye
\cite{11}. Then {\it theoretically} we can talk on the rough approximation for
the light quark. Indeed, since $m_q^{QCD} \ll \Lambda_{QCD}$ the light quark is
relativistic. Nevertheless, we introduce the system with a finite number of
degrees of freedom and an instantaneous interaction $V({\bf r})$. This fact is
a disadvantage because the confinement supposes the following: a) the
generation of sea around the light quark, i.e. the presence of infinite number
of gluons and quark-antiquark pairs, and b) the nonperturbative effects with
the correlation time $\tau_{QCD}\sim 1/\Lambda_{QCD}$, that is beyond the
potential approach. However, {\it phenomenologically} the introduction of
constituent mass $m_q^{NP}\sim \Lambda_{QCD}$ as a basic parameter determining
the interaction with the QCD condensates, allows us successfully to adjust the
nonrelativistic potential model with a high accuracy ($\delta M\approx 30 - 40$
MeV) by fitting the existing experimental data, that makes the approach to be
quite a reliable tool for the prediction of masses for the hadrons, containing
the heavy and light quarks.

As for the diquark $QQ'$, it is completely analogous to the heavy quarkonium
$Q\bar Q'$ except the very essential peculiarities.
\begin{enumerate}
\item
$(QQ')_{\bar 3_c}$ is a system with the nonzero color charge.
\item
For the quarks of the same flavor $Q=Q'$ it is necessary to take into account
the Pauli principle for the identical fermions.
\end{enumerate}
The second item turns out to forbid the sum of quark spins S=0 for the
symmetric, spatial parity P-even wave functions of diquark, $\Psi_d({\bf
r})$ (the orbital momentum equals $L_d= 2n$, where $n=0,1,2\ldots$ ), as well
as S=1 is forbidden for the anti-symmetric, Р-odd functions $\Psi_d({\bf r})$
(i.e. $L_d= 2n+1$). The nonzero color charge leads to two problems.

First, we cannot generally apply the confinement hypothesis on the form of
potential (an infinite growth of energy with the increase of the system size)
for the object under consideration. However, it is unpossible to imagine a
situation, when a big colored object with a size $r > 1/\Lambda_{QCD}$ has a
finite energy of self-action, and, to the same moment, it is confined inside a
white hadron (the singlet over SU(3)$_c$) with $r \sim 1/\Lambda_{QCD}$ due to
the interaction with another colored source. In the framework of well-justified
picture of the hadronic string, the tension of such string in the diquark with
the external leg inside the baryons is only two times less than in the
quark-antiquark pair inside the meson $q\bar q'$, and, hence, the energy of
diquark linearly grows with the increase of its size. So, the effect analogous
to the confinement of quarks takes place in the similar way. In the potential
models we can suppose that the quark binding appears due to the effective
single exchange by a colored object in the adjoint representation of SU(3)$_c$
(the sum of scalar and vector exchanges is usually taken). Then, the potentials
in the singlet ($q\bar q'$) and anti-triplet ($qq'$) states differ by the
factor of 1/2, that means the confining potential with the linear term in the
QCD-motivated models for the heavy diquark $(QQ')_{\bar 3_c}$. In the present
work we use the nonrelativistic model with the Buchm\" uleer--Tye potential for
the diquark, too.

Second, in the singlet color state $(Q\bar Q')$ there are separate
conservations of the summed spin S and the orbital momentum L, since the QCD
operators for the transitions between the levels determined by these quantum
numbers, are suppressed. Indeed, in the framework of multipole expansion in QCD
\cite{12}, the amplitudes of chromo-magnetic and chromo-electric dipole
transitions are suppressed by the inverse heavy quark mass, but in addition,
the major reason is provided by the following: a) the necessity to emit a white
object, i.e. at least two gluons, which results in the higher order in $1/m_Q$,
and b) the projection to a real phase space in a physical spectrum of massive
hadrons in contrast to the case of massless gluon. Furthermore, the probability
of a hybrid state, say, the octet sybsystem $(Q\bar Q')$ and the additional
gluon, i.e. the Fock state $|Q\bar Q^{\prime}_{8_c} g\rangle $, is suppressed
due to both the small size of system and the nonrelativistic motion of quarks
(for a more strict consideration see ref.\cite{14}). In the anti-triplet color
state, the emission of a soft nonperturbative gluon between the levels
determined by the spin $S_d$ and the orbital momentum $L_d$ of diquark, is not
forbidden, if there are no some other no-go rules or small order-parameters.
For the quarks of identical flavors inside the diquark, the Pauli principle
leads to that the transitions are possible only between the levels, which
either differ by the spin ($\Delta S_d= 1$) and the orbital momentum ($\Delta
L_d = 2n+1$), instantaniously, or belong to the same set of radial and orbital
excitations with $\Delta L_d = 2n$. Therefore, the transition amplitudes are
suppressed by a small recoil momentum of diquark in comparison with its mass.
The transition operator changing the diquark spin as well as its orbital
momentum, has the higher order of smallness because of either the additional
factor of $1/m_Q$ or the small size of diquark. These suppressions lead to the
existence of quazi-stable states with the quantum numbers of $S_d$ and $L_d$.
In the diquark composed by the quarks of different flavors, $bc$, the QCD
operators of dipole transitions with the single emission of soft gluon are not
forbidden, so that the lifetimes of levels can be comparable with the times for
the forming of bound states or with the inverse distances between the levels
themselves. Then, we cannot insist on the appearance of excitation system for
such the diquark with definite quantum numbers of the spin and orbital
momentum\footnote{In other words, the presence of gluon field inside the baryon
$\Xi_{bc}$ leads to the transitions between the states with the different
excitations of diquark, like $|bc\rangle \to |bcg\rangle $ with $\Delta S_d =1$
or $\Delta L_d=1$, which are not suppressed.}.

Thus, in the present work we explore the presence of two physical scales in the
form of factorization for the wave functions of the heavy diquark and light
constituent quark. So, in the framework of nonrelativistic quark model the
problem on the calculation of mass spectrum and characteristics of bound states
in the system of doubly heavy baryon is reduced to two standard problems on the
study of stationary levels of energy in the system of two bodies. After that,
we take into account the relativistic corrections dependent of the quark spins
in two subsystems under consideration. The natural boundary for the region of
stable states in the doubly heavy system can be assigned to the threshold
energy for the decay into a heavy baryon and a heavy meson. As was shown in
\cite{13}, the appearance of such threshold in different systems can be
provided by the existence of an universal characteristics in QCD, a critical
distance between the quarks. At distances greater than the critical separation,
the quark-gluon fields become unstable, i.e. the generation of valence
quark-antiquark pairs from the sea takes place. In other words, the hadronic
string having a length greater than the critical one, decay into the strings of
smaller sizes with a high probability close to unit. In the framework of
potential approach this effect can be taken into account by that we will
restrict the consideration of excited diquark levels by the region, wherein the
size of diquark is less than the critical distance, $r_{QQ'}< r_c \approx 1.4 -
1.5$ fm. Furthermore, the model with the isolated structure of diquark looks to
be reliable, just if the size of diquark is less than the distance to the light
quark $r_{QQ'}<r_l$.

The peculiarity of quark-diquark picture for the doubly heavy baryon is the
possibility of mixing between the states of higher diquark excitations,
possessing the different quantum numbers, because of the interaction with the
light quark. Then it is difficult to assign some definite quantum numbers to
such excitations. We will discuss the mechanism of this effect.

The paper is organized as follows. In Section II we describe a general
procedure for the calculation of masses for the doubly heavy baryons in
the framework of assumptions drawn above. We take into account the
spin-dependent corrections to the potential motivated in QCD. The results of
numerical estimates are presented in Section III, and, finally, our conclusions
are given in Section IV.

\section{Nonrelativistic potential model}

As we have mentioned in the Introduction, we solve the problem on the
calculation of mass spectra of baryons containing two heavy quarks,  in two
steps. First, we compute the energy levels of diquark. Second, we consider
the two body problem for the light quark interacting with the point-like
diquark having the mass obtained in the first step. In accordance with the
effective expansion of QCD in the inverse heavy quark mass, we separate two
stages of such the calculations. So, the nonrelativistic Schr\" odinger
equation with the model potential motivated by QCD, is solved numerically.
After that, the spin-dependent corrections are introduced as perturbations
suppressed by the quark masses.

\subsection{Potential}

We use the Buchm\" uller--Tye potential, which takes into account the
asymptotic freedon of QCD at short distances. So, the effective coupling
constant in the exchange by the octet color state between the quarks is
approximated by the QCD running coupling constant up to two-loop accuracy. At
long distances, there is the linear term of interaction energy, which leads to
the confinement. These two regimes are the limits for the effective
$\beta$-function by Gell-Mann--Low. It was given explicitly in \cite{11}. In
the anti-triplet quark state we introduce the factor of 1/2 because of the
color structure of bound quark-quark state. For the interaction of diquark with
the light constituent quark, the corresponding factor is equal to unit.

As was shown in \cite{15}, the nonperturbative constituent term
introduced into the mass of nonrelativistic quark, exactly coincides with the
additive constant, subtracted from the coulomb potential. We extract the masses
of heavy quarks by fitting the real spectra of charmonium and bottomonium,
\begin{equation}
m_c = 1.486\; \mbox{GeV},\quad m_b = 4.88\; \mbox{GeV},
\end{equation}
so that the mass of the level in the heavy quarkonium has been calculated as,
say, $M(с\bar c)=2 m_c+E$, where $E$ is the energy of stationary Schr\" odinger
equation with the model potential $V$. Then, we have supposed that the mass of
meson with a single heavy quark is equal to $M(Q\bar q) = m_Q + m_q + E$, and
$E=\langle T\rangle + \langle V-\delta V\rangle$, whereas the additive term in
the potential is introduced because the constituent mass of light quark is
determined as a part of interaction energy $\delta V = m_q$. In accordance with
fitting the masses of heavy mesons, we get $m_q = 0.385$ GeV. 

The results of calculations for the eigen-energies in the Schr\" odinger
equation with the Buchm\" uller--Tye potential are presented in Tabs. 1--3. The
characteristics of corresponding wave functions are
given in Tabs. 4--6.

We have checked that with a good accuracy {\it the binding energy
and the wave function of light quark practically do not depend on the flavors
of heavy quarks}. Indeed, large values of diquark masses give small
contributions into the reduced masses. This fact leads to small corrections to
the wave functions in the Schr\" odinger equation. So, for the states lying
below the threshold of doubly heavy baryon decay into the heavy baryon and
heavy meson, the energies of levels of light constituent quark are equal to
$$
E(1s) = 0.38\; {\rm GeV,}\;\;E(2s) = 1.09\; {\rm GeV,}\;\; E(2p) = 0.83\; {\rm
GeV,}
$$
where the energy has been defined as the sum of light quark constituent mass
and eigen-value of Schr\" odinger equation. In HQET the value of $\bar
\Lambda=E(1s)$ is generally introduced. Then we can draw a conclusion that our
estimate of $\bar \Lambda$ is in a good agreement with calculations in other
approaches \cite{10}. This fact confirms the reliability of such the
phenomenological predictions. For the light quark radial wave functions at the
origin we find
$$
R_{1s}(0) = 0.527\; {\rm GeV}^{3/2},\;\;
R_{2s}(0) = 0.278\; {\rm GeV}^{3/2},\;\;
R_{2p}'(0) = 0.127\; {\rm GeV}^{5/2}.\;\;
$$
The analogous characteristics of bound states of the $c$-quark
interacting with the $bb$-diquark, are equal to
$$
E(1s) = 1.42\; {\rm GeV,}\;\;E(2s) = 1.99\; {\rm GeV,}\;\; E(2p) = 1.84\; {\rm
GeV,}
$$
with the wave functions
$$
R_{1s}(0) = 1.41\; {\rm GeV}^{3/2},\;\;
R_{2s}(0) = 1.07\; {\rm GeV}^{3/2},\;\;
R_{2p}'(0) = 0.511\; {\rm GeV}^{5/2}.\;\;
$$
For the binding energy of strange constituent quark we add the current mass
$m_s\approx 100-150$ MeV.

\subsection{Spin-dependent corrections}

According to \cite{16}, we introduce the spin-dependent corrections causing
the splitting of $nL$-levels of diquark as well as in the system of light
constituent quark and diquark ($n=n_r+L+1$ is the principal number,
$n_r$ is the number of radial excitation, $L$ is the orbital momentum). For the
heavy diquark containing the identical quarks we have
\begin{eqnarray}
V_{SD}^{(d)}({\bf r}) &=& \frac{1}{2}\left(\frac{\bf L_d\cdot
S_d}{2m_Q^2}\right)
\left( -\frac{dV(r)}{rdr}+
\frac{8}{3}\alpha_s\frac{1}{r^3}\right)\nonumber \\
&& +\frac{2}{3}\alpha_s\frac{1}{m_Q^2}\frac{\bf L_d\cdot S_d}{r^3}+\frac{4}{3}
\alpha_s\frac{1}{3m_Q^2}{{\bf S}_{Q1}\cdot {\bf S}_{Q2}}[4\pi\delta({\bf r})]\\
&& -\frac{1}{3}\alpha_s\frac{1}{m_Q^2}\frac{1}{4{\bf L_d}^2 -3} [
6({\bf L_d\cdot S_d})^2+3({\bf L_d\cdot S_d})-2{\bf L_d}^2{\bf
S_d}^2]\frac{1}{r^3},
\nonumber     
\end{eqnarray}
where $\bf L_d,\; S_d$ are the orbital momentum in the diquark system and the
summed spin of quarks composing the diquark, respectively. Taking into
account the interaction with the light constituent quark gives
(${\bf S} = {\bf S_d}+{\bf S_l}$)
\begin{eqnarray}
V_{SD}^{(l)}({\bf r}) &=& \frac{1}{4}\left(\frac{\bf L\cdot S_d}{2m_Q^2}
+ \frac{2\bf L\cdot S_l}{2m_l^2}\right)
\left( -\frac{dV(r)}{rdr}+
\frac{8}{3}\alpha_s\frac{1}{r^3}\right)\nonumber \\
&& +\frac{1}{3}\alpha_s\frac{1}{m_Q m_l}\frac{(\bf L\cdot S_d + 
2L\cdot S_l)}{r^3}+ 
\frac{4}{3}\alpha_s\frac{1}{3m_Q m_l}{({\bf S_d}+{\bf L_d})\cdot {\bf S_l}}
[4\pi\delta({\bf r})]\\
&& -\frac{1}{3}\alpha_s\frac{1}{m_Q m_l}\frac{1}{4{\bf L}^2 -3} [
6({\bf L\cdot S})^2+3({\bf L\cdot S})-2{\bf L}^2{\bf S}^2\nonumber \\
&& -6({\bf L\cdot S_d})^2-3({\bf L\cdot S_d})+2{\bf L}^2{\bf S_d}^2]
\frac{1}{r^3}, \nonumber    
\end{eqnarray}
where the first term corresponds to the relativistic correction to the
effective scalar exchange, and other terms appear because of corrections to
the effective single-gluon exchange with the coupling constant $\alpha_s$. 

The value of effective parameter $\alpha_s$ can be determined in the following
way. The splitting in the $S$-wave heavy quarkonium $(Q_1\bar Q_2)$ is given by
the expression
\begin{equation}
\Delta M(ns) = \frac{8}{9}\alpha_s\frac{1}{m_1m_2}|R_{nS}(0)|^2,
\end{equation}
where $R_{nS}(r)$ is the radial wave function of quarkonium. From the
experimental data on the system of $c\bar c$ 
\begin{equation} 
\Delta M(1S,c\bar c) = 117\pm 2\; {\rm MeV,}
\end{equation} 
and $R_{1S}(0)$ calculated in the model, we can determine $\alpha_s(\Psi)$.
Let us take into account the dependence of this parameter on the reduced mass
of the system, $\mu $. In the framework of one-loop approximation for the
running coupling constant of QCD we have
\begin{equation} 
\alpha_s (p^2) = \frac{4\pi}{b\cdot\ln (p^2/\Lambda_{QCD}^2)}, 
\end{equation} 
whereas $b = 11 -2n_f/3$ and $n_f = 3$ at $p^2< m_c^2$. From the phenomenology
of potential models \cite{9} we know that the average kinetic energy of quarks
in the bound state practically does not depend on the flavors of quarks, and it
is given by the values
\begin{equation}
\langle T_{d}\rangle \approx 0.2\; {\rm GeV,}
\end{equation}
and
\begin{equation}
\langle T_{l}\rangle \approx 0.4\; {\rm GeV,}
\end{equation}
for the anti-triplet and singlet color states, correspondingly. Substituting
the definition of the nonrelativistic kinetic energy
\begin{equation}
\langle T\rangle  = \frac{\langle p^2\rangle }{2\mu },
\end{equation}
we get
\begin{equation}
\alpha_s(p^2) = \frac{4\pi}{b\cdot\ln (2\langle T\rangle \mu/\Lambda_{QCD}^2)},
\end{equation} 
whereas numerically $\Lambda_{QCD}\approx 113$ MeV.

For the identical quarks inside the diquark, the scheme of $LS$-coupling well
known for the corrections in the heavy quarkonium, is applicable. Otherwise,
for the interaction with the light quark we use the scheme of $jj$-coupling
(here, ${\bf LS_l}$ is diagonal at the given ${\bf J_l}$,  $({\bf J_l} =
{\bf L} + {\bf S_l}, {\bf J} = {\bf J_l} + {\bf \bar J})$, where $\bf J$
denotes the total spin of baryon, and $\bf\bar J$ is the total spin of
diquark, ${\bf\bar J}={\bf S_d}+{\bf L_d}$).

Then, to estimate various terms and mixings of states, we use
the transformations of bases (in what follows ${\bf S} = {\bf S_l} + {\bf\bar
J}$)
\begin{equation}
|J;J_l\rangle  = \sum_{S} (-1)^{(\bar J+S_l+L+J)}\sqrt {(2S+1)(2J_l+1)}
\left\{\begin{array}{ccc} \bar J & S_l & S \\
                        L & J & J_l \end{array}\right\}|J;S\rangle ,
\end{equation}
and 
\begin{equation}
|J;J_l\rangle  = \sum_{J_d} (-1)^{(\bar J+S_l+L+J)}\sqrt {(2J_d+1)(2J_l+1)}
\left\{\begin{array}{ccc} \bar J & L & J_d \\
                        S_l & J & J_l \end{array}\right\}|J;J_d\rangle .
\end{equation}
Thus, we have defined the procedure of calculations for the mass spectra of
doubly heavy baryons. This procedure leads to results presented in the next
section.

\section{Numerical results}
In this Section we present the results on the mass spectra with account for the
spin-dependent splitting of levels. As we have clarified in the Introduction,
the doubly heavy baryons with identical heavy quarks allow quite a reliable
interpretation in terms of diquark quantum numbers (the summed spin and the
orbital momentum). Dealing with the excitations of $bc$-diquark, we show the
results on the spin-dependent splitting of the ground 1S-state, since the
emission of soft gluon breaks the simple classification of levels for the
higher excitations of such diquark.

For the doubly heavy baryons, the quark-diquark model of bound states obviously
leads to the most reliable results for the system with the larger mass of heavy
quark, i.e. for $\Xi_{bb}$.

\subsection{$\Xi_{bb}$ baryons}
For the quantum numbers of levels, we use the notations
$n_d L_d n_l l_l$, i.e. we show the value of principal quantum number of
diquark, its orbital momentum by a capital letter and the principal quantum
number for the excitations of light quark and its orbital momentum by a
lower-case letter. We denote the shift of level by $\Delta^{(J)}$ as dependent
on the total spin of baryon $J$. So, for $1S2p$ we have
\begin{equation}
\Delta^{(\frac{5}{2})} = 10.3~{\rm MeV}.
\end{equation}
The states with the total spin $J = \frac{3}{2}$ (or $\frac{1}{2}$), can have
different values of $J_l$, and, hence, they have a nonzero mixing, when we
perform the calculations in the perturbation theory built over the states with
the definite total momentum $J_l$ of the light constituent quark. For $J
=\frac{3}{2}$, the
mixing matrix equals
\begin{equation}
\left(\begin{array}{cc} -3.0 & -0.5 \\
                        -0.5 & 11.4 \end{array}\right)~{\rm MeV},
\end{equation}
so that the mixing practically can be neglected, and the level shifts are
determined by the values
\begin{eqnarray}
\Delta^{\prime(\frac{3}{2})} =\lambda_1^{\prime } &=& -3.0~{\rm MeV},\\
\Delta^{(\frac{3}{2})} =\lambda_1 &=& 11.4~{\rm MeV}.\nonumber
\end{eqnarray}
For $J=\frac{1}{2}$, the mixing matrix has the form
\begin{equation}
\left(\begin{array}{cc} -5.7 & -17.8 \\
                        -17.8 & -14.9 \end{array}\right)~{\rm MeV},
\end{equation}
with the eigen-vectors given by 
\begin{eqnarray}
|1S2p(\frac{1}{2}^{\prime })\rangle  &=& 0.790|J_l=\frac{3}{2}\rangle
-0.613|J_l=\frac{1}{2}\rangle ,\\
|1S2p(\frac{1}{2})\rangle  &=& 0.613|J_l=\frac{3}{2}\rangle
+0.790|J_l=\frac{1}{2}\rangle ,
\nonumber
\end{eqnarray}
and the eigen-values equal
\begin{eqnarray}
\Delta^{\prime(\frac{1}{2})} =\lambda_2^{\prime } &=& 8.1~{\rm MeV},\\
\Delta^{(\frac{1}{2})} =\lambda_2 &=& -28.7~{\rm MeV}.\nonumber
\end{eqnarray}
For the $2S2p$-level, the corresponding quantities are equal to
\begin{equation}
\Delta^{(\frac{5}{2})} = 10.3~{\rm MeV},
\end{equation}
and for $J =\frac{3}{2}$, the mixing matrix is equal to
\begin{equation}
\left(\begin{array}{cc} -3.6 & -0.5 \\
                        -0.5 & 12.4 \end{array}\right)~{\rm MeV},
\end{equation}
so that
\begin{eqnarray}
\Delta^{\prime(\frac{3}{2})} =\lambda_1^{\prime } &=& -3.6~{\rm MeV},\\
\Delta^{(\frac{3}{2})} =\lambda_1 &=& 12.4~{\rm MeV}.\nonumber
\end{eqnarray}
For $J=\frac{1}{2}$, the matrix has the form
\begin{equation}
\left(\begin{array}{cc} -6.1 & -17.6 \\
                        -17.6 & -13.5 \end{array}\right)~{\rm MeV,} 
\end{equation}
with the eigen-vectors
\begin{eqnarray}
|1S2p(\frac{1}{2}^{\prime })\rangle  &=& 0.776|J_l=\frac{3}{2}\rangle
-0.631|J_l=\frac{1}{2}\rangle ,\\
|1S2p(\frac{1}{2})\rangle  &=& 0.631|J_l=\frac{3}{2}\rangle
+0.776|J_l=\frac{1}{2}\rangle ,
\nonumber
\end{eqnarray}
possessing the eigen-values
\begin{eqnarray}
\Delta^{\prime(\frac{1}{2})} =\lambda_2^{\prime } &=& 8.2~{\rm MeV},\\
\Delta^{(\frac{1}{2})} =\lambda_2 &=& -27.8~{\rm MeV}.\nonumber
\end{eqnarray}
We can straightforwardly see, that the difference between the wave functions
as caused by the different masses of diquark subsystem, is unessential.

The splitting of diquark, $\Delta^{(J_d)}$, has the form
\begin{eqnarray}
&3D1s:&\nonumber \\
\Delta^{(3)} &=& -0.06~{\rm MeV},\nonumber\\
\Delta^{(2)} &=& 0.2~{\rm MeV},\\
\Delta^{(1)} &=& -0.2~{\rm MeV}.\nonumber
\end{eqnarray}
\begin{eqnarray}
&4D1s:&\nonumber \\
\Delta^{(3)} &=& -2.6~{\rm MeV},\nonumber\\
\Delta^{(2)} &=& -0.8~{\rm MeV},\\
\Delta^{(1)} &=& -4.6~{\rm MeV}.\nonumber
\end{eqnarray}
\begin{eqnarray}
&5D1s:&\nonumber \\
\Delta^{(3)} &=& 2.6~{\rm MeV},\nonumber\\
\Delta^{(2)} &=& -0.9~{\rm MeV},\\
\Delta^{(1)} &=& -4.7~{\rm MeV}.\nonumber
\end{eqnarray}
\begin{eqnarray}
&5G1s:&\nonumber \\
\Delta^{(5)} &=& -0.3~{\rm MeV},\nonumber\\
\Delta^{(4)} &=& 0.3~{\rm MeV},\nonumber\\
\Delta^{(3)} &=& 1.1~{\rm MeV},\\
\Delta^{(2)} &=& 1.7~{\rm MeV},\nonumber\\
\Delta^{(1)} &=& 2.0~{\rm MeV}.\nonumber
\end{eqnarray}
\begin{eqnarray}
&6G1s:&\nonumber \\
\Delta^{(5)} &=& 3.2~{\rm MeV},\nonumber\\
\Delta^{(4)} &=& -0.5~{\rm MeV},\nonumber\\
\Delta^{(3)} &=& -4.4~{\rm MeV},\\
\Delta^{(2)} &=& -7.9~{\rm MeV},\nonumber\\
\Delta^{(1)} &=& -10.5~{\rm MeV}.\nonumber
\end{eqnarray}
Such the corrections are unessential up to the current accuracy of method
($\delta M\approx 30 - 40$ MeV). They can be neglected for the excitations,
whose sizes are less than the distance to the light quark, i.e. for
the diquarks with small values of principal quantum number.

For the hyper-fine spin-spin splitting in the system of quark-diquark, we have
\begin{equation}
\Delta_{h.f.}^{(l)} = \frac{2}{9}\bigg[J(J+1)-\bar J(\bar J +1 ) -
\frac{3}{4}\bigg]
\alpha_s(2\mu T)\frac{1}{m_cm_l} |R_l(0)|^2,
\end{equation}
where $R_l(0)$ is the radial wave function at the origin for the light
constituent quark, and for the analogous shift of diquark level, we find
\begin{equation}
\Delta_{h.f.}^{(d) }= \frac{1}{9}
\alpha_s(2\mu T)\frac{1}{m_c^2} |R_d(0)|^2.
\end{equation}

The mass spectra of $\Xi_{bb}^{+}$ and $\Xi_{bb}^{0}$ baryons are shown in
Fig.1 and in Tab. 7, wherein we restrict ourselves by the presentation of S-,
P- and D-wave levels.

We can see in Fig.1, that the most reliable predictions are the masses of
baryons $1S1s\; (J^P=3/2^+,\; 1/2^+)$, $2P1s\; (J^P=3/2^-,\; 1/2^-)$ and
$3D1s\; (J^P=7/2^+,\ldots 1/2^+)$. The $2P1s$-level is quazi-stable, because
the transition into the ground state requires the instantaneous change of both
the orbital momentum and the summed spin of quarks inside the diquark. The
analogous kind of transitions seems to be the transition between the states of
ortho- and para-hydrogen in the molecule of $H_2$. This transition take place
in a non-homogeneous external field due to the magnetic moments of other
molecules. For the transition of $2P1s\to 1S1s$, the role of such the external
field is played by the non-homogeneous chromo-magnetic field of the light
quark. The corresponding perturbation has the form
\begin{eqnarray}
\delta V & \sim & \frac{1}{m_Q} [{\bf S}_1\cdot {\bf H}_1 + {\bf S}_2\cdot {\bf
H}_2 - ({\bf S}_1+{\bf S}_2)\cdot \langle{\bf H}\rangle ] \nonumber\\
& = & \frac{1}{2m_Q}({\bf \nabla\cdot r_d})\; ({\bf S}_1-{\bf S}_2)\cdot {\bf
H} \sim \frac{1}{m_Q}\frac{{\bf r_l}\cdot {\bf r_d}}{m_q r_l^5}\; ({\bf
S}_1-{\bf S}_2)\cdot{\bf J_l}\; f(r_l),\nonumber
\end{eqnarray}
where $f(r_l)$ is a dimensionless nonperturbative function depending on the
distance between of the light quark and diquark. The $\delta V$
operator changes the orbital momentum of  light quark, too. It results in the
mixing between the states with the same values of $J^P$. If the splitting is
not small (for instance, $2P1s - 1S2p$, where $\Delta E\sim \Lambda_{QCD}$),
then the mixing is suppressed as $\delta V/\Delta E \sim \frac{1}{m_Q m_q}
\frac{r_d}{r_l^4} \frac{1}{\Delta E} \ll 1$. Since the admixture of $1S2p$ in
the $2P1s$-state is low, the  $2P1s$-levels are quazi-stable, i.e. their
hadronic transitions into the ground state with the emission of $\pi$-mesons 
are suppressed as we have derived, though an additional suppression is given by
a small value of phase space. Therefore, we have to expect the presence of
narrow resonances in the mass spectra of pairs $\Xi_{bb} \pi$, as they are
produced in the decays of quazi-stable states with $J^P=3/2^-,\; 1/2^-$. The
experimental observation of such levels could straightforwardly confirm the
existence of diquark excitations and provide the information on the character
of dependency in $f(r_l)$, i.e. on the non-homogeneous chromo-magnetic field in
the nonperturbative region.

Sure, the $3D1s\; J^P=7/2^+,\; 5/2^+$ states are also quazi-stable, since in
the framework of multipole expansion in QCD they transform into the ground
state due to the quadru-pole emission of gluon (the E2-transition with the
hadronization $gq\to q' \pi$).

As for the higher excitations, the $3P1s$-states are close to the $1S2p$-levels
with $J^P=3/2^-,\; 1/2^-$, so that the operators changing both the
orbital momentum of diquark and its spin, can lead to the essential mixing with
an amplitude $\delta V_{nn'}/\Delta E_{nn'}\sim 1$, despite of suppression by
the inverse heavy quark mass and small size of diquark. We are sure that
the mixing slightly shifts the masses of states. The most important effect is
a large admixture of $1S2p$ in $3P1s$. It makes the state to be unstable
because of the transition into the ground $1S1s$-state with the emission of
gluon (the E1-transition). This transition leads to decays with the emission of
$\pi$-mesons\footnote{Remember, that the $\Xi_{QQ'}$-baryons are the
iso-dublets.}. 

The level $1S2p\; J^P=5/2^-$ has the definite quantum numbers of diquark and
light quark motion, because there are no levels with the same values of $J^P$
in its vicinity. However, its width of transition into the ground state and
$\pi$-meson is not suppressed and seems to be large, $\Gamma\sim 100$ MeV.

The following transitions take place
$$
{\frac{3}{2}}^- \to {\frac{3}{2}}^+ \pi \;\; {\rm in~ S-wave,}
$$
$$
{\frac{3}{2}}^- \to \frac{1^+}{2} \pi \;\; {\rm in~ D-wave,}
$$
$$
{\frac{1}{2}}^- \to {\frac{3}{2}}^+ \pi \;\; {\rm in~ D-wave,}
$$
$$
{\frac{1}{2}}^- \to {\frac{1}{2}}^+ \pi \;\; {\rm in~ S-wave.}
$$
The D-wave transitions are suppressed by the ratio of low recoil momentum to
the mass of baryon.

The width of state $J^P=3/2^+$ is completely determined by the
radiative electromagnetic M1-transition into the basic $J^P=1/2^+$ state.

\subsection{$\Xi_{cc}$ baryons}
The calculation procedure described above leads to the results for the
doubly charmed baryons as presented below.

For $1S2p$, the splitting is equal to
\begin{equation}
\Delta^{(\frac{5}{2})} = 17.4~{\rm MeV}.
\end{equation}
For $J =\frac{3}{2}$, the mixing is determined by the matrix
\begin{equation}
\left(\begin{array}{cc} 4.3 & -1.7 \\
                        -1.7 & 7.8 \end{array}\right)~{\rm MeV,} 
\end{equation}
so that the eigen-vectors
\begin{eqnarray}
|1S2p(\frac{3}{2}^{\prime })\rangle  &=& 0.986|J_l=\frac{3}{2}\rangle
+0.164|J_l=\frac{1}{2}\rangle ,\\
|1S2p(\frac{3}{2})\rangle  &=& -0.164|J_l=\frac{3}{2}\rangle
+0.986|J_l=\frac{1}{2}\rangle ,
\nonumber
\end{eqnarray}
have the eigen-values
\begin{eqnarray}
\Delta^{\prime(\frac{3}{2})} =\lambda_1^{\prime } &=& 3.6~{\rm MeV},\\
\Delta^{(\frac{3}{2})} =\lambda_1 &=& 8.5~{\rm MeV}.\nonumber
\end{eqnarray}
For $J=\frac{1}{2}$, the mixing matrix equals
\begin{equation}
\left(\begin{array}{cc} -3.6 & -55.0 \\
                        -55.0 & -73.0 \end{array}\right)~{\rm MeV,} 
\end{equation}
where the vectors
\begin{eqnarray}
|1S2p(\frac{1}{2}^{\prime })\rangle  &=& 0.957|J_l=\frac{3}{2}\rangle
-0.291|J_l=\frac{1}{2}\rangle ,\\
|1S2p(\frac{1}{2})\rangle  &=& 0.291|J_l=\frac{3}{2}\rangle
+0.957|J_l=\frac{1}{2}\rangle ,
\nonumber
\end{eqnarray}
have the eigen-values
\begin{eqnarray}
\Delta^{\prime(\frac{1}{2})} =\lambda_2^{\prime } &=& 26.8~{\rm MeV},\\
\Delta^{(\frac{1}{2})} =\lambda_2 &=& -103.3~{\rm MeV}.\nonumber
\end{eqnarray}
The splitting of $3D$ diquark level is given by
\begin{eqnarray}
\Delta^{(3)} &=& -3.02~{\rm MeV},\nonumber\\
\Delta^{(2)} &=& 2.19~{\rm MeV},\\
\Delta^{(1)} &=& 3.39~{\rm MeV}.\nonumber
\end{eqnarray}
Further, we have to take into account the hyper-fine spin-spin corrections in
the quark-diquark system.

For the $1S$- and $2S$-wave levels of diquark, the shifts of vector states are
equal to
\begin{eqnarray}
\Delta (1S) &=& 6.3~{\rm MeV},\nonumber\\
\Delta (2S) &=& 4.6~{\rm MeV}.\nonumber
\end{eqnarray}

The mass spectra of the $\Xi_{cc}^{++}$ and $\Xi_{cc}^{+}$ baryons are
presented in Fig.2 and Tab.8.

\subsection{$\Xi_{bc}$ baryons}
As we have already mentioned in the Introduction, the heavy diquark composed of
the quarks of different flavors, turns out to be unstable under the
emission of soft gluons. So, in the Fock state of doubly heavy baryon, there is
a sizable nonperturbative admixture of configurations
including the gluons and diquark with the various values of its spin $S_d$ and
orbital momentum $L_d$
$$
|B_{bcq}\rangle = O_B|bc_{\bar 3_c}^{S_d,L_d},q\rangle+
H_1|bc_{\bar 3_c}^{S_d\pm 1,L_d},g,q\rangle+
H_2|bc_{\bar 3_c}^{S_d,L_d\pm 1},g,q\rangle+\ldots ,
$$
whereas the amplitudes of $H_1$, $H_2$ are not suppressed with respect to
$O_B$. In the heavy quarkonium, the analogous operators for the octet-color
states are suppressed by the probability of emission by the nonrelativistic
quarks inside a small volume determined by the size of singlet-color system of
heavy quark and antiquark. In the baryonic system under consideration, a soft
gluon is restricted only by the ordinary scale of confinement, and, hence,
there is no suppression.

We suppose that the calculations of masses for the excited $\Xi_{bc}$ baryons
are not so justified in the given scheme. Therefore, we present only the result
for the ground state with $J^P=1/2^+$
$$
M_{\Xi_{bc}^{\prime}} = 6.85\; {\rm GeV,}\;\;\;
M_{\Xi_{bc}} = 6.82\; {\rm GeV,}
$$
whereas for the vector diquark we have assumed that the spin-dependent
splitting due to the interaction with the light quark is determined by the
standard contact coupling of magnetic moments for the point-like systems. The
picture for the baryon levels is shown in Fig.3 with no account for the
spin-dependent perturbations suppressed by the heavy quark masses.

\subsection{The doubly heavy baryons with the strangeness $\Omega_{QQ'}$.}

In the leading approximation, we suppose that the wave functions and the
excitation energies of strange quark in the field of doubly heavy diquark
repeat the characteristics for the analogous baryons containing the ordinary
quarks $u,\; d$. Therefore, the level system of baryons $\Omega_{QQ'}$
reproduces that of $\Xi_{QQ'}$ up to an additive shift of the masses by the
value of current mass of strange quark, $m_s \approx M(D_s)-M(D) \approx
M(B_s)-M(B) \approx 0.1$ GeV.

Further, we suppose that the spin-spin splitting of $2P1s$ and $3D1s$ levels of
$\Omega_{QQ'}$ is 20-30\% less than  in $\Xi_{QQ'}$ (the factor of
$m_{u,d}/m_s$). As for the $1S2p$-level, the procedure described above can be
applied. So, for $\Omega_{bb}$, the matrix of mixing for the states with the
different values of total momentum $J_l$ practically can be assigned to be
diagonal. This fact means that the following term of perturbation is dominant:
$$
\frac{1}{4}\left( \frac{2\bf L\cdot S_l}{2m_l^2}\right)
\left( -\frac{dV(r)}{rdr}+ \frac{8}{3}\alpha_s\frac{1}{r^3}\right).
$$
Therefore, we can think that the splitting of $1S2p$ is determined by the
factor of $m_{u,d}^2/m_s^2$ with respect to the splitting of corresponding
$\Xi_{bb}$, i.e. it is 40 \% less than in $\Xi_{bb}$. Hence, the splitting is
very small.

For the baryon $\Omega_{сс}$, the factor of $m_s/m_c$ is not small. Hence,
for $1S2p$, the mixing matrix is not diagonal, so that the arrangement of
$1S2p$ states of $\Omega_{сс}$ can be slightly different from that of
$\Xi_{сс}$.

The following peculiarity of $\Omega_{QQ'}$ is of great interest: the low-lying
$S$- and $P$-excitations of diquark are stable. Indeed, even after taking into
account the mixing of levels, a gluon emission makes a hadronization into the
$K$-meson (the transitions of $\Omega_{QQ'}\to \Xi_{QQ'}+K$), while a single
emission of $\pi$-meson is forbidden because of the conservation of iso-spin
and strangeness. The hadronic transitions with kaons are forbidden because of
insufficient splitting between the masses of $\Omega_{QQ'}$ and $\Xi_{QQ'}$.
The decays with the emission of pion pairs belonging to the iso-singlet state,
are suppressed by a small phase space or even forbidden. Thus, the radiative
electromagnetic transitions into the ground state are the dominant modes of
decays for the low-lying excitations of $\Omega_{QQ'}$.

\subsection{$\Omega_{bbc}$ baryons}
In the framework of quark-diquark picture, we can build the model for the
baryons containing three heavy quarks, $bbc$. However, as we estimate, the size
of diquark turns out to be comparable with the average distance to the charmed
quark. So, the model assumption on the compact heavy diquark cannot be quite
accurate for the calculations of mass levels in this case. The spin-dependent
forces are negligibly small inside the diquark, as we have already pointed out
above. The spin-spin splitting of vector diquark interacting with the charmed
quark, is given by
$$
\Delta (1s) =  33\; {\rm MeV,}\;\;
\Delta (2s) =  18\; {\rm MeV.}\;\;
$$
For $1S2p$, the level shifts are small. So, for the state $J^P=1/2^-$ we have
to add the correction of $- 33$ MeV. For the  $3D1s$-state the splitting is
determined by the spin-spin interaction. The characteristics of excitations for
the charmed quark in the model with the potential by Buchm\" uller--Tye have
been presented above. Finally, we obtain the picture of $\Omega_{bbc}$ levels
presented in Fig.4 and Tab.9.

Further, the excitations of ground $\Omega_{bbc}^0$ state can strongly mix with
large amplitudes because of small splittings between the levels, but they have
small shifts of masses. This effect takes place for $3P1s$ -- $1S2p$ with $J^P=
1/2^-,\; 3/2^-$, and for $2S1s$ -- $3D1s$ with $J^P= 1/2^+,\; 3/2^+$. We
suppose the prediction to be quite reliable for the states of $1S1s$ with
$J^P=1/2^+,\; 3/2^+$, $1S2p$ with $J^P=5/2^-$ and $3D1s$ with $J^P=5/2^+,\;
7/2^+$.  For these excitations, we might definitely predict the widths of their
radiative electromagnetic transitions into the ground state in the framework of
multipole expansion in QCD. The widths for the transitions will be essentially
determined by the amplitudes of admixtures, which have a strong model
dependence. Therefore, the experimental study of electromagnetic transitions in
the family of $\Omega_{bbc}^0$ baryons could provide a valuable information on
the mechanism of mixing between the different levels in the baryonic systems.
The electromagnetic transitions combined with the emission of pion pairs, if
not forbidden by the phase space, saturate the total widths of excited
$\Omega_{bbc}^0$ levels. The characteristic value of total width is about
$\Gamma \sim 10 - 100$ keV, in the order of magnitude.

Thus, the system of $\Omega_{bbc}^0$ can be characterized by a large number
of narrow quazi-stable states.

\section{Conclusion}
In this paper we have calculated the spectroscopic characteristics of baryons
containing two heavy quarks, in the model with the quark-diquark factorization
of wave functions. We have explored the nonrelativistic model of constituent
quarks with the potential by Buchm\" uller--Tye. The region of applicability of
such the approximations has been pointed out.

We have taken into account the spin-dependent relativistic corrections to the
potential in the subsystems of diquark and light quark-diquark. Below the
threshold of decay into the heavy baryon and heavy meson, we have found the
system of excited bound states, which are quazi-stable under the hadronic
transitions into the ground state. We have considered the physical reasons for
the quazi-stability taking place for the baryons with two identical quarks. In
accordance with the Pauli principle, the operators responsible for the
hadronic decays and the mixing between the levels, are suppressed by the
inverse heavy quark mass and the small size of diquark. This suppression is
caused by the necessity of instantaneous change in both the spin and the
orbital momentum of compact diquark. In the baryonic systems with two heavy
quark and the strange quark, the quazi-stability of diquark excitations is
provided by the absense of transitions with the emission of both a single kaon
and a single pion. These transitions are forbidden because of small splitting
between the levels and the conservation of the iso-spin and strangeness.

The characteristics of wave functions can be used in calculations of
cross-sections for the doubly heavy baryons in the framework of quark-diquark
approximation.

This work is in part supported by the Russian Foundation for Basic Research,
grants 99-02-16558 and 00-15-96645. The work of A.I.Onishchenko was supported,
in part, by International Center of Fundamental Physics in Moscow,
International Science Foundation, and INTAS-RFBR-95I1300 grants.


\newpage
\begin{table}[th]
\begin{center}
\begin{tabular}{|c|c|c|c|c|c|}
diquark level  & mass (GeV) &  $\langle r^2\rangle ^{1/2}$ (fm) & 
diquark level  & mass (GeV) &  $\langle r^2\rangle ^{1/2}$ (fm) \\ 
\hline 
1S & 9.74 & 0.33 & 2P & 9.95 & 0.54\\
2S & 10.02 & 0.69 & 3P & 10.15 & 0.86\\ 
3S & 10.22 & 1.06 & 4P & 10.31 & 1.14\\
4S & 10.37 & 1.26 & 5P & 10.45 & 1.39\\
5S & 10.50 & 1.50 & 6P & 10.58 & 1.61\\
3D & 10.08 & 0.72 & 4D & 10.25 & 1.01 \\
5D & 10.39 & 1.28 & 6D & 10.53 & 1.51 \\
4F & 10.19 & 0.87 & 5F & 10.34 & 1.15 \\
6F & 10.47 & 1.40 & 5G & 10.28 & 1.01 \\
6G & 10.42 & 1.28 & 6M & 10.37 & 1.15\\
\end{tabular} 
\end{center} 
\caption{The spectrum of $bb$-diquark levels without spin-dependent
splittings: masses and mean- squared radii.}
\end{table}

\begin{table}[th]
\begin{center}
\begin{tabular}{|c|c|c|c|c|c|}
diquark level & mass (GeV) &  $\langle r^2\rangle ^{1/2}$ (fm) & 
diquark level & mass (GeV) &  $\langle r^2\rangle ^{1/2}$ (fm) \\ 
\hline 
1S & 6.48 & 0.48 & 3P & 6.93 & 1.16\\
2S & 6.79 & 0.95 & 4P & 7.13 & 1.51\\ 
3S & 7.01 & 1.33 & 3D & 6.85 & 0.96\\
2P & 6.69 & 0.74 & 4D & 7.05 & 1.35\\
4F & 6.97 & 1.16 & 5F & 7.16 & 1.52\\
5G & 7.09 & 1.34 & 6H & 7.19 & 1.50\\
\end{tabular} 
\end{center} 
\caption{The spectrum of $bc$-diquark levels without spin-dependent
splittings: masses and mean- squared radii.}
\end{table}

\begin{table}[th]
\begin{center}
\begin{tabular}{|c|c|c|c|c|c|}
diquark level & mass (GeV) &  $\langle r^2\rangle ^{1/2}$ (fm) & 
diquark level & mass (GeV) &  $\langle r^2\rangle ^{1/2}$ (fm) \\ 
\hline 
1S & 3.16 & 0.58 & 3P & 3.66 & 1.36\\
2S & 3.50 & 1.12 & 4P & 3.90 & 1.86\\ 
3S & 3.76 & 1.58 & 3D & 3.56 & 1.13\\
2P & 3.39 & 0.88 & 4D & 3.80 & 1.59\\
\end{tabular} 
\end{center} 
\caption{The spectrum of $cc$-diquark levels without spin-dependent
splittings: masses and mean-squared radii.}
\end{table}

\begin{table}[th]
\begin{center}
\begin{tabular}{|c|c|c|c|}
nL  & $R_{d(ns)}(0)$  & nL & 
$R_{d(np)}'(0)$  \\
\hline 
1S & 1.346 & 2P & 0.479  \\ 
2S & 1.027 & 3P & 0.539  \\
3S & 0.782 & 4P & 0.585  \\
4S & 0.681 & 5P & 0.343  \\
\end{tabular}
\end{center}
\caption{The characteristics of radial wave function for the $bb$-diquark:
$R_{d(ns)}(0)$ (GeV$^{3/2}$), $R_{d(np)}^{'} (0)$ (GeV$^{5/2}$).}
\end{table} 

\begin{table}[th]
\begin{center}
\begin{tabular}{|c|c|c|c|}
nL  & $R_{d(ns)}(0)$  & nL & 
$R_{d(np)}'(0)$  \\
\hline 
1S & 0.726 & 2P & 0.202  \\ 
2S & 0.601 & 3P & 0.240  \\
3S & 0.561 & 4P &    \\
\end{tabular}
\end{center}
\caption{The characteristics of radial wave function for the $bc$-diquark:
$R_{d(ns)}(0)$ (GeV$^{3/2}$), $R_{d(np)}^{'} (0)$ (GeV$^{5/2}$).}
\end{table} 

\begin{table}[th]
\begin{center}
\begin{tabular}{|c|c|c|c|}
nL  & $R_{d(ns)}(0)$  & nL & 
$R_{d(np)}'(0)$  \\
\hline 
1S & 0.530 & 2P & 0.128  \\ 
2S & 0.452 & 3P & 0.158  \\
\end{tabular}
\end{center}
\caption{The characteristics of radial wave function for the $cc$-diquark:
$R_{d(ns)}(0)$ (GeV$^{3/2}$), $R_{d(np)}^{'} (0)$ (GeV$^{5/2}$).}
\end{table} 

\begin{table}[th]
\begin{center}
\begin{tabular}{|p{40mm}|c|p{40mm}|c|}
$(n_d L_d n_l L_l)$, $J^{P}$ & mass ({\rm GeV})
 & $(n_d L_d n_l L_l)$, $J^{P}$ & mass ({\rm GeV})  \\
\hline
(1S 1s)$1/2^{+}$ & 10.093 & (3P 1s)$1/2^{-}$    & 10.493 \\
(1S 1s)$3/2^{+}$ & 10.133  & (3D 1s)$5/2^{\prime +}$ & 10.497 \\
(2P 1s)$1/2^{-}$ & 10.310  & (3D 1s)$7/2^{+}$   & 10.510 \\
(2P 1s)$3/2^{-}$ & 10.343 & (3P 1s)$3/2^{-}$   & 10.533 \\
(2S 1s)$1/2^{+}$ & 10.373 & (1S 2p)$1/2^{-}$ & 10.541  \\
(2S 1s)$3/2^{+}$ & 10.413 & (1S 2p)$3/2^{-}$ & 10.567 \\
(3D 1s)$5/2^{+}$ & 10.416  & (1S 2p)$1/2^{\prime -}$ & 10.578 \\
(3D 1s)$3/2^{\prime +}$ & 10.430 & (1S 2p)$5/2^{-}$  & 10.580 \\
(3D 1s)$1/2^{+}$ & 10.463 & (1S 2p)$3/2^{\prime -}$ & 10.581  \\
(3D 1s)$3/2^{+}$ & 10.483 & (3S 1s)$1/2^{+}$   &  10.563 \\
\end{tabular}
\end{center}
\caption{The mass spectrum of $\Xi_{bb}^{-}$ and $\Xi_{bb}^{0}$ baryons.}
\end{table}

\begin{table}[th]
\begin{center}
\begin{tabular}{|p{40mm}|c|p{40mm}|c|}
$(n_d L_d n_l L_l)$, $J^{P}$ & mass ({\rm GeV})
 & $(n_d L_d n_l L_l)$, $J^{P}$ & mass ({\rm GeV})  \\
\hline
(1S 1s)$1/2^{+}$ & 3.478 & (3P 1s)$1/2^{-}$    & 3.972 \\
(1S 1s)$3/2^{+}$ & 3.61  & (3D 1s)$3/2^{\prime +}$ & 4.007 \\
(2P 1s)$1/2^{-}$ & 3.702  & (1S 2p)$3/2^{\prime -}$ & 4.034  \\
(3D 1s)$5/2^{+}$ & 3.781  & (1S 2p)$3/2^{-}$ & 4.039 \\
(2S 1s)$1/2^{+}$ & 3.812 & (1S 2p)$5/2^{-}$  & 4.047\\
(3D 1s)$3/2^{+}$ & 3.83 & (3D 1s)$5/2^{\prime +}$ & 4.05 \\
(2P 1s)$3/2^{-}$ & 3.834 & (1S 2p)$1/2^{\prime -}$ & 4.052 \\
(3D 1s)$1/2^{+}$ & 3.875 & (3S 1s)$1/2^{+}$   & 4.072\\
(1S 2p)$1/2^{-}$ & 3.927  & (3D 1s)$7/2^{+}$   & 4.089 \\
(2S 1s)$3/2^{+}$ & 3.944 & (3P 1s)$3/2^{-}$   & 4.104 \\
\end{tabular}
\end{center}
\caption{The mass spectrum of $\Xi_{cc}^{++}$ and $\Xi_{cc}^{+}$ baryons.}
\end{table}

\begin{table}[t]
\begin{center}
\begin{tabular}{|p{40mm}|c|p{40mm}|c|}
$(n_d L_d n_l L_l)$, $J^{P}$ & mass ({\rm GeV})
 & $(n_d L_d n_l L_l)$, $J^{P}$ & mass ({\rm GeV})  \\
\hline
(1S 1s)$1/2^{+}$ & 11.12 & (3D 1s)$3/2^{\prime +}$ & 11.52 \\
(1S 1s)$3/2^{+}$ & 11.18  & (3D 1s)$5/2^{\prime +}$ & 11.54 \\
(2P 1s)$1/2^{-}$ & 11.33  & (1S 2p)$1/2^{-}$ & 11.55 \\
(2P 1s)$3/2^{-}$ & 11.39  & (3D 1s)$7/2^{+}$   & 11.56 \\
(2S 1s)$1/2^{+}$ & 11.40 & (1S 2p)$3/2^{\prime -}$ & 11.58 \\
(3D 1s)$5/2^{+}$ & 11.42  & (1S 2p)$3/2^{-}$ & 11.58 \\
(3D 1s)$3/2^{+}$ & 11.44 & (1S 2p)$1/2^{\prime -}$ & 11.59 \\
(3D 1s)$1/2^{+}$ & 11.46 & (1S 2p)$5/2^{-}$  & 11.59\\
(2S 1s)$3/2^{+}$ & 11.46  & (3P 1s)$3/2^{-}$   & 11.59 \\
(3P 1s)$1/2^{-}$ & 11.52 & (3S 1s)$1/2^{+}$   & 11.62 \\
\end{tabular}
\end{center}
\caption{The mass spectrum of $\Omega_{bbc}^0$ baryons.}
\end{table}

\newpage

\begin{figure}[th]
\hspace*{-4cm}
\epsfxsize=21cm \epsfbox{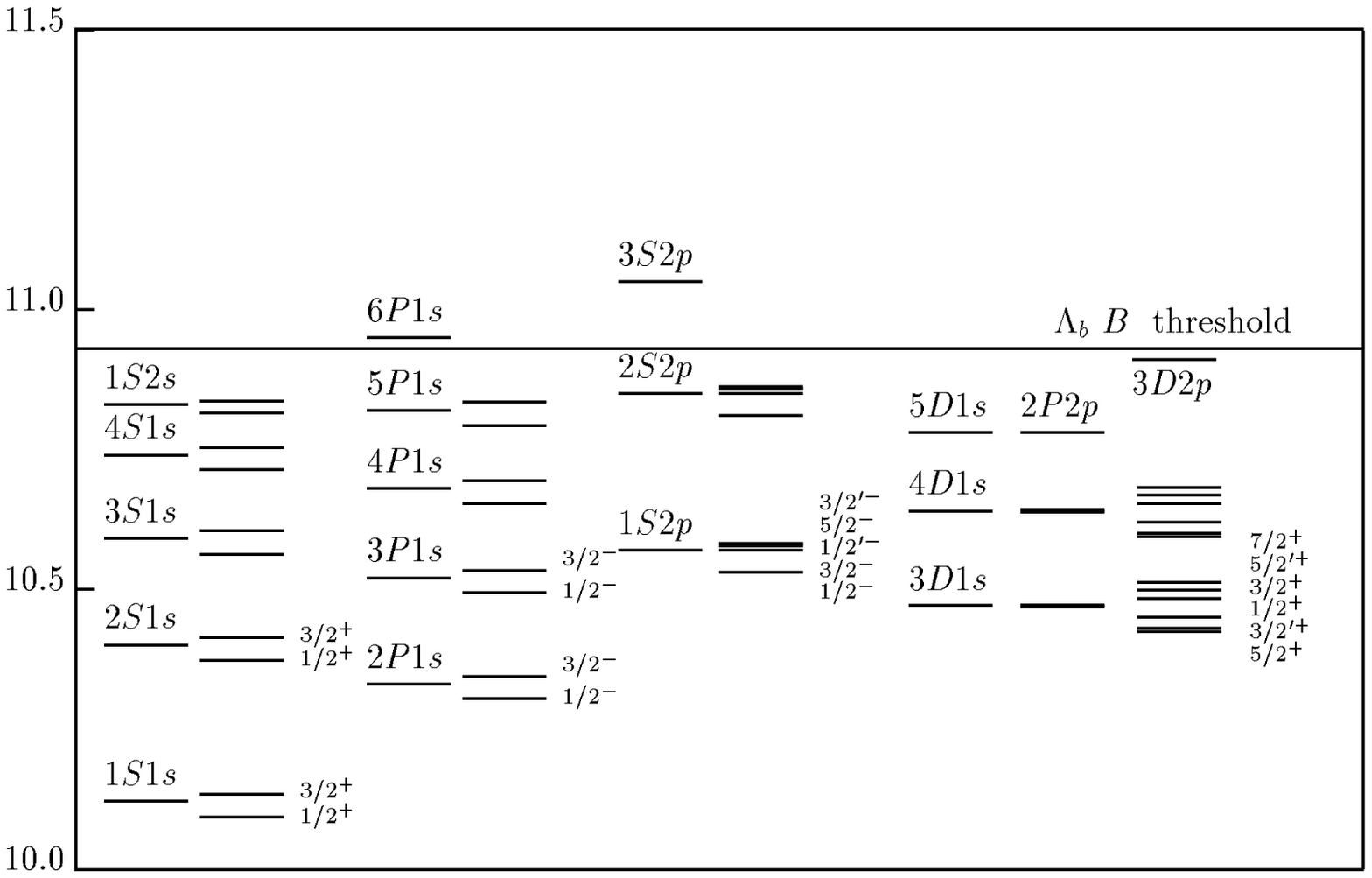}

\caption{The spectrum of baryons, containing two $b$-quarks: $\Xi_{bb}^{-}$ and
$\Xi_{bb}^{0}$, with account for the spin-dependent splittings of low-lying
excitations. The masses are given in GeV.}
\label{pic-bb}
\end{figure}

\begin{figure}[th]
\hspace*{-4cm}
\epsfxsize=21cm \epsfbox{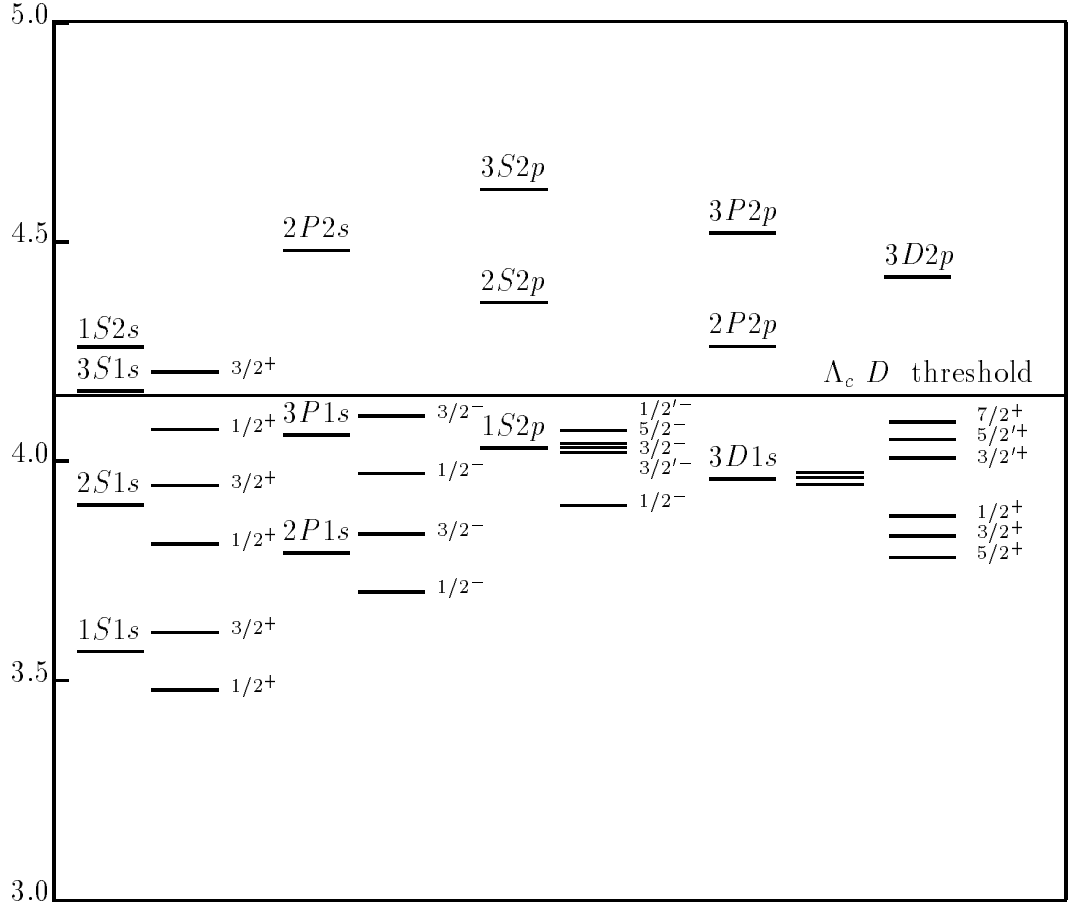}

\caption{The spectrum of $\Xi_{cc}^{++}$ and $\Xi_{cc}^{+}$ baryons. The masses
are given in GeV.}
\label{pic1}
\end{figure}

\begin{figure}[t]
\hspace*{-4cm}
\epsfxsize=21cm \epsfbox{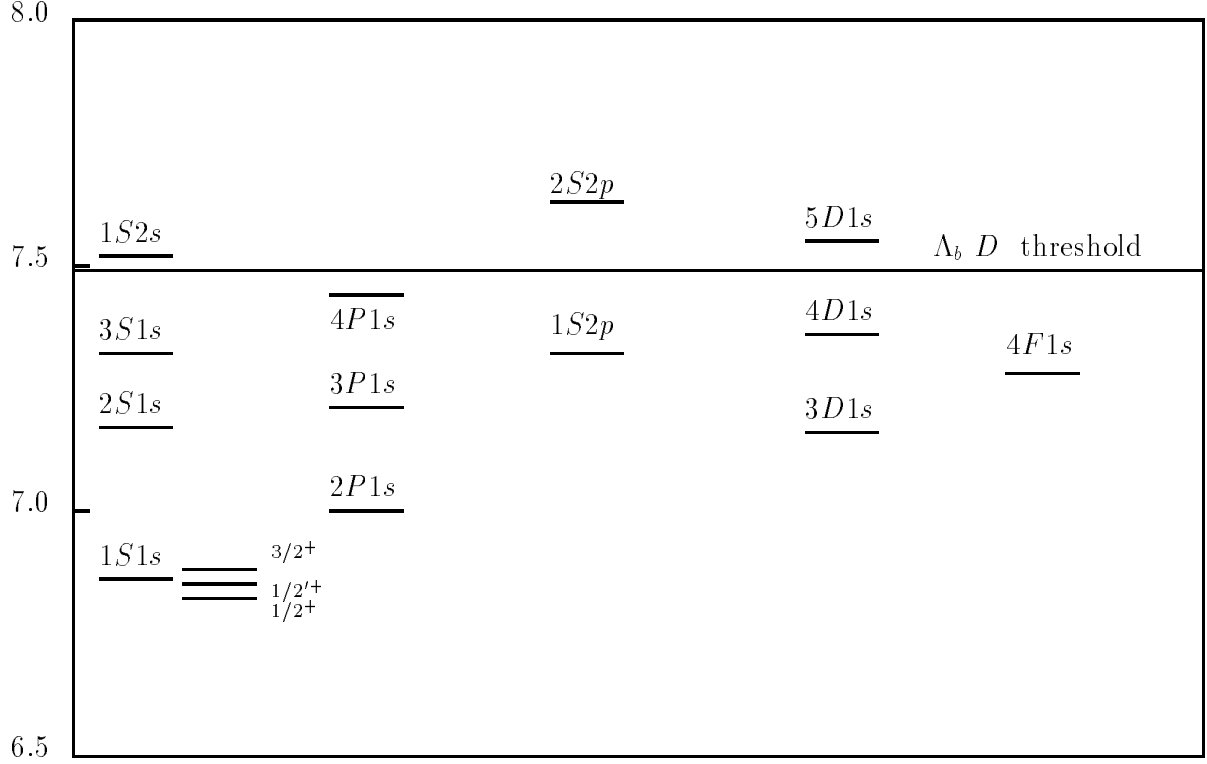}

\caption{The spectrum of $\Xi_{bc}^{+}$ and $\Xi_{bc}^{0}$ baryons without the
splittings of higher excitations. The masses are given in GeV.}
\label{pic-bc}
\end{figure}

\newpage
\begin{figure}[t]
\hspace*{-4cm}
\epsfxsize=21cm \epsfbox{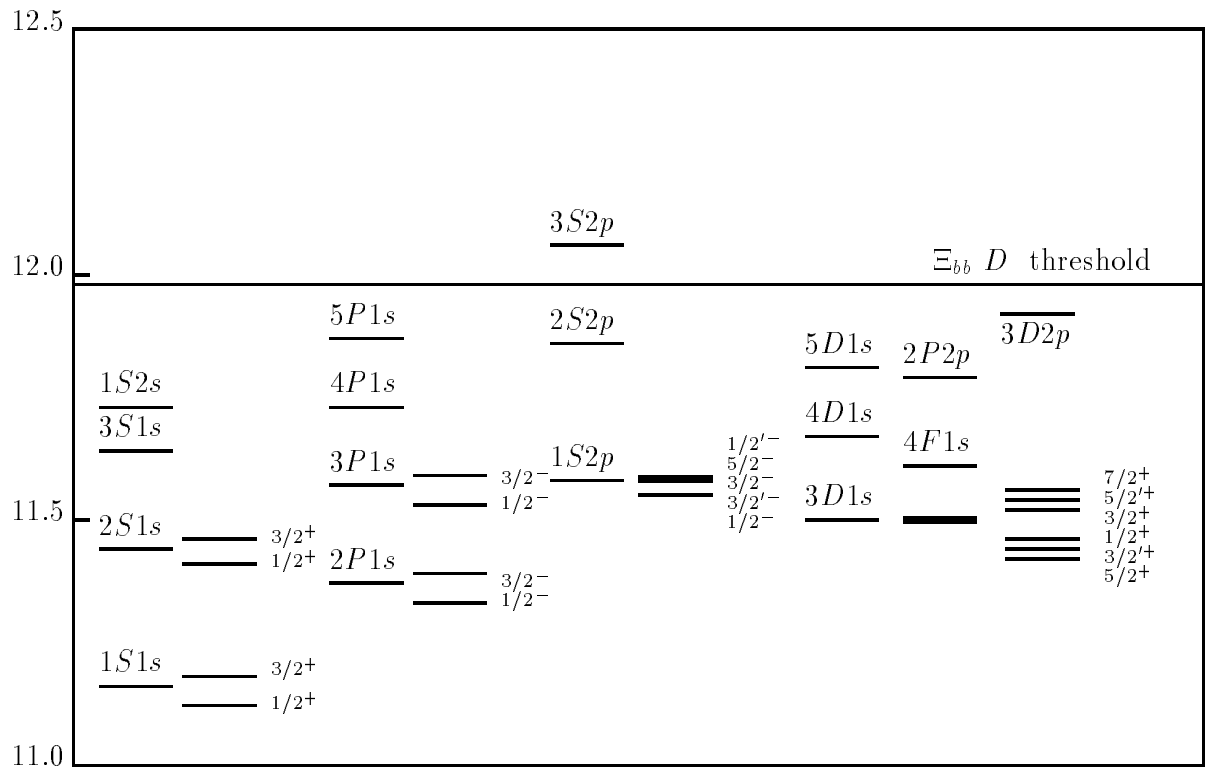}

\caption{The spectrum of $\Omega_{bbc}^0$ baryons with account for the
spin-dependent splittings for the low-lying excitations. Masses are given in
GeV.}
\label{pic-bbc}
\end{figure}

\end{document}